\documentclass[prl,aps,twocolumn,showpacs,superscriptaddress,floatfix,amsmath,amssymb,nofootinbib]{revtex4}
\usepackage{amsfonts}
\usepackage{graphicx} 
\newcommand{\beq}{\begin{equation}}
\newcommand{\eeq}{\end{equation}}

\newcommand{\beqa}{\begin{eqnarray}}
\newcommand{\eeqa}{\end{eqnarray}}

\begin{document} 
\title{Explanation and observability of diffraction in time}
\author{E. Torrontegui}
\affiliation{Departamento de Qu\'{\i}mica F\'{\i}sica, Universidad del Pa\'{\i}s Vasco - Euskal Herriko Unibertsitatea, 
Apdo. 644, Bilbao, Spain}
\affiliation{Max Planck Institute f\"ur Physik Complexer Systeme, N\"othnitzer Stra$\beta$e 38, D-01187 Dresden, Germany}
\author{J. Mu\~noz}
\affiliation{Departamento de Qu\'{\i}mica F\'{\i}sica, Universidad del Pa\'{\i}s Vasco - Euskal Herriko Unibertsitatea, 
Apdo. 644, Bilbao, Spain}
\author{Yue Ban}
\affiliation{Departamento de Qu\'{\i}mica F\'{\i}sica, Universidad del Pa\'{\i}s Vasco - Euskal Herriko Unibertsitatea, 
Apdo. 644, Bilbao, Spain}
\author{J. G. Muga}
\affiliation{Departamento de Qu\'{\i}mica F\'{\i}sica, Universidad del Pa\'{\i}s Vasco - Euskal Herriko Unibertsitatea, 
Apdo. 644, Bilbao, Spain}
\affiliation{Max Planck Institute f\"ur Physik Complexer Systeme, N\"othnitzer Stra$\beta$e 38, D-01187 Dresden, Germany}
\begin{abstract}
Diffraction in time (DIT) is a fundamental phenomenon in quantum dynamics due to time-dependent obstacles and slits. It is formally analogous to diffraction of light, and is expected to play an increasing role to design coherent matter wave 
sources, as in the atom laser, to analyze time-of-flight information 
and emission from ultrafast pulsed excitations, and in applications of coherent matter waves in integrated atom-optical circuits. 
We demonstrate that DIT emerges robustly in quantum waves emitted by an exponentially decaying source and provide a simple explanation of the phenomenon, as an interference of two characteristic velocities. This allows for its controllability and optimization.      
\end{abstract}  	
\maketitle
Diffraction in time is a fundamental quantum dynamical effect first studied by  Moshinsky \cite{Mos}. 1D matter waves released through a time-modulated
aperture or encountering a time-dependent obstacle (for 2D and 3D cases see \cite{Zeilinger,Godoy3D}) show temporal quantum penumbras and interference patterns similar to the diffraction of light behind spatial slits and obstacles.  
Understanding and controlling DIT is more and more relevant as a result of the increasing manipulability of coherent matter waves, in particular in ultracold atomic gases and/or with ultrashort laser pulses. DIT will affect, for example, the intended applications of atom lasers \cite{dCMM}, dynamics of matter waves emitted by ultrashort laser excitations \cite{Paulus}, matter-wave circuits \cite{Pfau}, and time-of-flight techniques \cite{Golub}. DIT may also lead to temporal versions of diffractometers, grating spectrometry, and holography.  

The original and most studied setting for DIT is the Moshinsky shutter (MS). It consists in a sudden release, by opening a shutter, of a semi-infinite plane-wave beam characterized by a ``carrier''  velocity. The particle density as a function of time at an observation point is formally analogous to spatial Fresnel diffraction by a sharp edge \cite{Mos}. If 
the shutter, when closed, has reflecting amplitude $R=1$, the same results are obtained from a point source with a sharp onset and constant emission thereafter
\cite{review}.
Many works have applied and modified MS to study different quantum transients and, adding a potential, resonance scattering, buildup and decay, and tunneling dynamics, see e.g. \cite{MML,Gas} and reviews in  \cite{Kleber,review}. Experimentally,  
a DIT oscillatory pattern was first observed by Dalibard and coworkers with cold atoms falling by gravity and bouncing off a mirror made of  evanescent light that could be switched on and off \cite{Dalibard}.
DIT through a related time-energy relation has been observed for cold neutron experiments too \cite{Golub}. Interferences from two time slits and time-analogues of diffraction from a grating have been described for cold atoms \cite{Dalibard,BECS} and ionizing atoms with ultrashort laser pulses \cite{Paulus}. There are also  
analogs of the original MS in the field of coherent 
transients due to frequency-chirped weak lasers \cite{atoms1}.

DIT may be suppressed or averaged out by apodization, noise and decoherence, or unsharp carrier velocity distributions \cite{Godoy,review}, so 
the observability of MS-DIT with matter waves has been considered a difficult task \cite{review}. We shall see, however, that the effect is rather robust and occurs quite generally in waves emitted by an exponentially decaying resonance.   

A second problematic aspect of MS-DIT is the lack of a simple and intuitive understanding of the phenomenon. The usual, geometrical ``explanation'' in terms of a Cornu spiral \cite{Mos,Zeilinger,review} does not provide a simple physical picture although  
some insight is gained by its construction via Fresnel time-zones and Huygens principle, as in spatial diffraction
\cite{Zeilinger}. Also, an attempt was made in \cite{Wigner} to seek an explanation in terms of the Wigner distribution but, as recognized by the authors, the interpretation of the results remained ambiguous due to the lack of positivity of the Wigner function.   

\begin{figure}[t]
\label{compa}
\begin{center}
\includegraphics[height=4cm,angle=0]{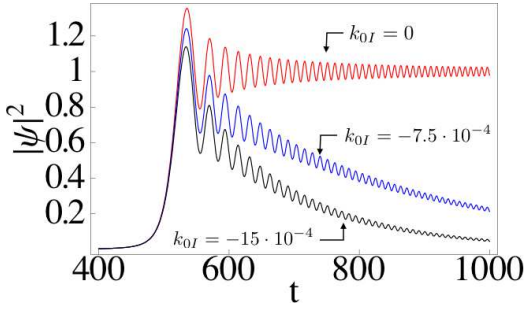}
\end{center}
\caption{\label{f0}
(Color online)
Unnormalized density versus time at $x=1000$ 
for a constant or exponentially decaying source.
} 
\end{figure}

In this paper we address the observability and interpretation of DIT, as well as some important consequences.
They are linked to each other since a simple physical explanation of DIT will also provide the key to observe and control it. The starting point is the realization that systems that decay exponentially
due to a resonance, such as cold atoms in magnetic or optical traps \cite{Mark} escaping from their initial confinement, 
may show DIT at a distance from the trap. The density or flux oscillations will be  identified as an interference effect characterized quantitatively with a simple  analytical model \cite{torr}. We shall thus be able to predict and design  optimal conditions for its observability,  
and treat on the same footing the standard constant emission after a sharp onset, and the exponentially decaying source, by modifying continuously the imaginary part of the emission pole.   
Figure \ref{f0}, discussed later in more detail, 
shows the unnormalized density at an observation point away from the  source. The upper curve corresponds to ordinary MS-DIT oscillations. 
%
The amplitude of the oscillations at the observation point decreases with time, and their frequency depends on time, tending to a constant. 
The other two curves correspond to exponentially decaying sources with different lifetimes. The oscillations are essentially the same as in the standard MS,  
modulated by the exponential decay.

{\it{The exponentially decaying source model.}}
We shall use a model that captures the essence of resonance decay from a trap and describes analytically the external wave function without the complications and peculiarities of particular confinements \cite{torr}. We adopt the same
notation as in \cite{torr} with dimensionless position $x$, time $t$, and wave function $\psi$ 
obeying formally a Schr\"odinger equation for a particle of mass $1/2$ and $\hbar=1$,
%
$i{\partial\psi(x,t)}/{\partial t}=-{\partial^2\psi(x,t)}/{\partial x^2}$. 
The unit of length corresponds to the inverse of the 
real part of the carrier wavenumber and the unit of time to the carrier period divided by $2\pi$.   
The complex dimensionless wavenumber $k_0=k_{0R}+ik_{0I}$ and frequency of the carrier $\omega_0=\omega_{0R}+i\omega_{0I}$,  obey the dispersion relation 
%
%
$\omega_{0}=k_{0}^{2}=(1+ik_{0I})^2$,   
%
so $\omega_{0R}=1-k_{0I}^2$ and $\omega_{0I}=2k_{0I}$, with $k_{0I}<0$ and $k_{0R}=1$.  
The exact unnormalized solution to the Schr\"odinger equation for the free particle subjected to the source boundary condition
%
$\psi(0,t)=e^{-i\omega_0 t}\Theta(t),\;\;\;\;\omega_{0R}>0,\, \omega_{0I}<0$, 
%
can be constructed by a superposition of plane waves. The resulting integral 
can be expressed in terms of known functions, 
%
$\psi(x,t)=\frac{1}{2}e^{ik_s^2 t}\left[w(-u_0^{(+)})+w(-u_0^{(-)})\right],
$
%
where $w(z):=e^{-z^{2}}{\rm{erfc}}(-i z)$,
%
$u_0^{(\pm)}=\pm(1+i)\sqrt{{t}/{2}}k_0(1\mp\tau/t)$,
and
%
$k_{s}={x}/{2t}$, $\tau={x}/{2k_0}$
%
are a ``saddle point'' wave number and a complex characteristic time.  
For an observation point $x$, the saddle velocity is time dependent,  $v_s=2k_s=x/t$.
{}Figure \ref{f0} shows the unnormalized density $|\psi(x,t)|^2$ for different $k_{0I}$ to illustrate the essential continuity of oscillation phenomena when varying $k_{0I}$. 
If one particle is emitted, the normalized
wave function is  
%
$\tilde \psi(x,t)=[{\int_{0}^{\infty}\!\!dt\,J(0,t)}]^{-1/2}\psi(x,t)$, 
%
where $J(x,t)$ is the dimensionless flux.  
%
\begin{figure}[t]
\begin{center}
\includegraphics[height=4cm,angle=0]{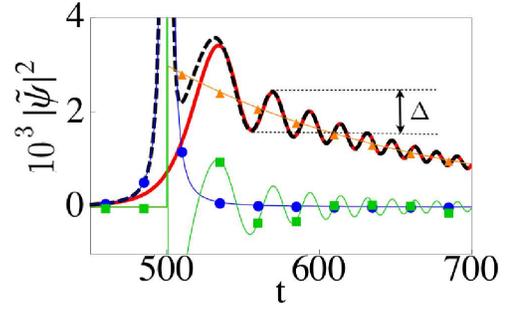}
\end{center}
\caption{\label{f1}
(Color online) 
Normalized probability densities versus time for $k_0=1-0.0015i$ at $x=1000$. 
Exact (red solid line), approximate (Eq. (\ref{approx2}), black dashed line), saddle term (blue circles), 
pole term (orange triangles),
and interference term (Eq. (\ref{int}), green squares).} 
\end{figure}

{\it{The essence of DIT.}}
The wavefunction $\psi$,  
for times shorter and larger than $|\tau|$ \cite{torr}, can be accurately approximated by contributions from the two critical points of its defining integral, saddle and pole,    
%
$\psi(x,t)=\psi_s(x,t)+\psi_{0}(x,t)\Theta[\mbox{Im}(u_{0}^{(+)})]$, 
%
where
$\psi_s(x,t) = 
({2t}/{\pi})^{1/2}{\tau e^{ik_s^2t}}/[(i-1)k_0(t^2-\tau^2)]$, 
and
$\psi_{0}=
e^{-i\omega_0 t}e^{ik_0 x}$.
As a result  of contour deformation along the steepest descent 
path from the saddle, the pole term contributes from the time 
when $\mbox{Im}(u^{(+)}_0)=0$, $t_c=x/[2(1+k_{0I})]$,
and decays exponentially thereafter. In a pictorial, classical association \cite{torrcl}, 
the particle arriving at $(x,t)$ with velocity $v_0=2$ must have been released at a time $x/v_0$
from the source which emits particles exponentially.   
The saddle velocity $x/t$ is the one required for a classical  particle
released from $(0,0)$ to arrive at $(x,t)$.
Saddle trajectories may thus be pictured as the result of a burst or ``big bang'' emerging from the source with all possible velocities
at $t=0$. These classical pictures are useful 
but, unlike long-time deviations from exponential
decay \cite{torrcl}, DIT cannot be explained with them alone. It is a quantum interference phenomenon as shown by 
the structure of the unnormalized density   
\beqa
\label{approx2}
|\psi(x,t)|^2&=&|\psi_s(x,t)|^2+|\psi_0(x,t)|^2\Theta[\mbox{Im}(u_{0}^{+})] \nonumber \\
&+&2\mbox{Re}[\psi_s(x,t)\psi_{0}^{*}(x,t)]\Theta[\mbox{Im}(u_{0}^{+})],
\label{den}
\eeqa
where the asterisk denotes complex conjugation. 
The interference term is 
\beqa
&&2\mbox{Re}[\psi_s\psi_{0}^{*}]=\sqrt{\frac{t}{\pi}}
\frac{2\beta(x,t)}{16|\omega_{0}|^2t^4+x^4-8t^2x^2 \omega_{0R}}
\nonumber\\
&\times&\left[
({8\omega_{0R}xt^2-2x^3})
\cos \phi  
+{8\omega_{0I}xt^2}
\sin \phi\right],
\label{int} 
\eeqa
%
where $\phi(x,t)=(\omega_{0R}+k_{s}^{2})t-x-\frac{3\pi}{4}$, 
and $\beta(x,t)=e^{\omega_{0I}t-k_{0I}x}$. 
Light does not show DIT in vacuum because
there is no dispersion and no interference of this kind.   

Figure \ref{f1} shows the agreement at times larger and shorter than $|\tau|$ between exact and  approximate wave functions.
The pole and saddle terms separately do not oscillate in time, whereas the interference term, Eq. (\ref{int}), reproduces accurately the characteristic oscillations of DIT.

{\it{Characterization and observability of DIT.}}
%
%
%
\begin{figure}[t]
 \begin{center}
\includegraphics[height=4cm,angle=0]{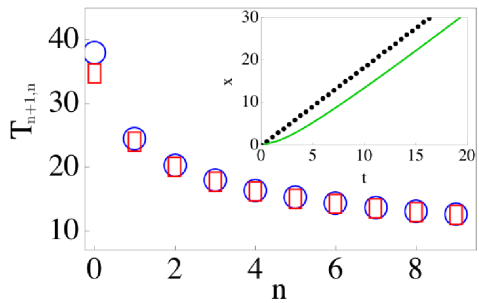}
 \end{center}
\caption{\label{f2}
(Color online) Time intervals $T_{n+1}-T_n$ between two consecutive maxima:  
exact (circles),
and approximation from Eq. (\ref{tn}) (squares).
Same parameters as in Fig. \ref{f1}. The symbols size is to help the eye, not related to errors.  
In the inset: nonlinear position of the first maximum versus time for $k_{0I}=-0.08$ (solid line) and onset of the pole term $x_c=2(1+k_{0I})t$ (dotted line).} 
\end{figure}
The frequency of the DIT oscillation
depends on the interference of the saddle and pole 
frequencies $k_s^2$ and $\omega_{0R}$ and,  
as $k_s$ depends on time,   
the DIT oscillation period is not
constant.
{}From Eq. (\ref{int}) we can infer the position of the {\itshape n}-th maximum. For $|k_{0I}|\ll 1$, the $\sin\phi$ term of Eq. (\ref{int}) tends to vanish 
so the DIT oscillations are essentially due to the $\cos\phi$ term.  
The maxima correspond to $\phi(x,T_n)=2n\pi$ at times 
\beq T_n=\frac{(3+8n)\pi+4x+\sqrt{[(3+8n)\pi+4x]^2-16\omega_{0R}x^2}}{8\omega_{0R}},
\label{tn}
\eeq
where $n=0, 1, 2, \dots$ ($n=0$ is for the principal maximum). The interval $T_{n+1,n}\equiv T_{n+1}-T_{n}$ between two consecutive maxima   
is in good agreement with the exact, numerically calculated period, see Fig. \ref{f2}. The small discrepancy at $n=0$ can be attributed to the dependence on time of the factors multiplying $\cos\phi$ and the proximity of $|\tau|$. 

For large times the period of the DIT oscillations tends
to the carrier period, 
%
$\displaystyle\lim_{n \to\infty}{T_{n+1,n}}={2\pi}/{\omega_{0R}}$. 
%
The amplitude of the oscillations decays relatively slowly compared to the pole term, as $e^{\omega_{0I}t}t^{-3/2}$, see Eq. (\ref{int}), but exponentially faster than the saddle term.   
\begin{figure}[t]
\begin{center}
\includegraphics[height=4cm,angle=0]{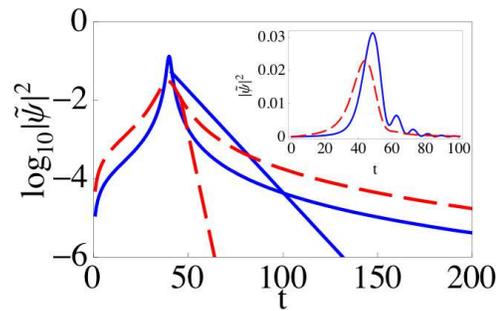}
\end{center}
\caption{\label{sp}
(Color online)  
Saddle and pole terms for $k_{0I}=-0.03$ (solid blue line)
and  $k_{0I}=-0.13$ (long-dashed red line). $x=80$. The inset shows the corresponding exact densities. The shorter lifetime suppresses DIT. 
} 
\end{figure}

\begin{figure}[t]
\begin{center}
\includegraphics[height=3.5cm,angle=0]{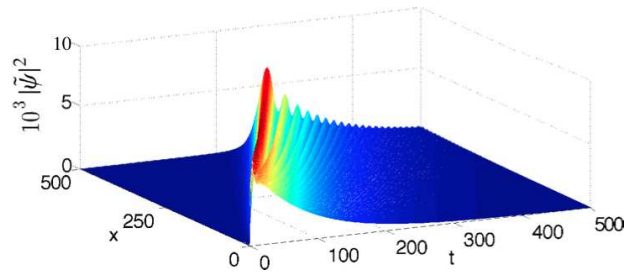}
\end{center}
\caption{\label{3d}
(Color online) Density $|\tilde{\psi}(x,t)|^2$ for $k_{0I}=-0.003$
showing the transition from pure exponential decay 
to a DIT pattern.}  
\end{figure}
According to Eq. (\ref{tn}), $T_0$ is not a linear function of $x$.  
For example, in the limit $k_{0I}\to 0$  
the motion of the first maximum is described by $x_{0}=2T_0-\sqrt{3T_0\pi}$.
Thus, even though an asymptotic velocity may be defined, $2(1+k_{0I})$ in the general case, see the inset of Fig. \ref{f2}, there is no oblique asymptote for this function. Therefore a naive linear extrapolation back to the origin at some large distance fails to provide the instant of the source onset. In other words, the times in which the tangents to $x_0(t)$ cut $x_0=0$ have no definite limit, in spite of the well defined asymptotic velocity. This is an example of the importance of DIT to correct simple classical-dynamical extrapolation from asymptotic wave features to extract emission
characteristics, as practiced e.g. in the analysis of ionization by ultrashort laser pulses \cite{Yue}.
         
In our dimensionless description two factors affect the 
visibility of the DIT pattern: the observation position $x$ and the lifetime.
Figure \ref{sp} 
shows the moduli of the logarithm of the pole and saddle densities for two 
different lifetimes. The pole term is a semi-infinite straight line which begins when the pole 
is crossed by the steepest descent path passing along the saddle in the complex momentum plane, at $t_c$; the saddle term shows a maximum near $|\tau|$ and decays from there slowly. There may be up to two intersections of the two terms, one near the arrival of the main front,
and one at a long time  that marks the transition to post-exponential decay \cite{torr}. 
When the saddle and pole terms are similar or close enough the interference oscillations appear. The interference region of our interest here is the one following the main front 
because it relates by continuity to 
ordinary MS-DIT in the limit
$k_{0I}\to 0$; it is also much more easily observable than the oscillations at large times because of the magnitude of the amplitudes.  

The oscillations are evidently
not present at the source $x=0$, and will be small at small distances, $x\lesssim 1$,  because of the rapid decay and separation from the pole term of the saddle term in these conditions, see Fig. \ref{3d}. The saddle term beyond the main front arrival increases with $x$ \cite{torr}. In the opposite extreme of very large $x$, it eventually dominates entirely and stays above the pole term at all times, 
suppressing DIT and even exponential decay \cite{torr}. Between these two extreme scenarios there is an ample range of $x$ for which DIT is prominent. The slope of the pole term also plays a role. For  larger values (smaller lifetimes), pole and saddle contributions separate more rapidly leading to fewer visible DIT oscillations       
which may actually disappear for small enough lifetimes.

To estimate the domain where some oscillations are seen before the decay is too strong we may solve 
$T_{1,0}<N\tau_{0}$ for a small $N$, where  $\tau_0=1/4|k_{0I}|$
is the lifetime.  This gives an explicit but lengthy expression. For $N=5$
and in the $k_{0I}\rightarrow 0$ limit, 
$x\lesssim 
30 \tau_0^2$.
 
{}From the previous discussion it might seem that a very long lifetime is always preferable to attain DIT. Nevertheless long lifetimes  
also imply a weaker signal because of the normalization. The consequence of opposite tendencies is  
an optimal lifetime-position point.  
A good measure of the visibility of DIT is 
the difference $\Delta$ between the second maximum and the previous  
minimum of the 
normalized probability density, see Fig. \ref{f1}.
The optimal parameters are found to be    
$k_{0I}=-0.03$, $x=60$. 
%

%
%
\begin{figure}[t]
\begin{center}
\includegraphics[height=4cm,angle=0]{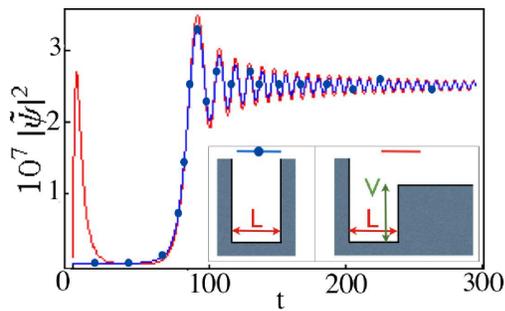}
\end{center}
\caption{\label{e1}
(Color online) Stability of DIT for decay from $U\delta(x)$ (Winter's model). 
The initial states 
are the ground states of the two 
wells in the inset.  At $t=0$ the right wall, at $x=0$, is substituted by the delta. 
$L=3.14$, $x=157.05$, 
$U=161.35$, $V=202.72$. 
} 
\end{figure}

{\it Model independence of the results.}
%
%
%
%
We have described the close connection between DIT and resonance decay. 
DIT will occur when contributions from different resonances
are well separated, which generally requires narrow and/or strong confinement. 
DIT does not depend on the specific properties of the model used so far.  
We have checked the robustness of DIT from exponential decay 
explicitly with several additional models.   
Winter's decay model \cite{Winter61} describes the decay 
of the ground state of the square well between $-L$ and $0$
when the right infinite wall is substituted by a 
$U\delta(x)$ potential. The wave function outside the trap  
tends to the source model wavefunction
for large $U$ \cite{capi}. Moreover DIT does not depend dramatically on the
strict confinement of the initial wave function on a finite domain.
To show this we
have calculated the decay of the ground state of a well with a
finite right wall once this wall is substituted by the delta.  
This produces a different fast forerunner at $x$, but  
the part associated with the dominant, lowest energy resonance remains essentially 
stable showing DIT as for the infinite wall, see Fig. \ref{e1}.    
Moreover we have observed the same stability for finite-width barriers. 
DIT also survives a smooth source onset 
\cite{dCMM}, and again may be observed after the passage of some 
onset-dependent transients. 
As for the effect of collisions, in the mean field regime
DIT is enhanced for attractive interactions \cite{review}.          
   
Let us finally point out the possibility to observe DIT in periodic structures 
\cite{MML} such as optical lattices, or other physical systems that realize a tight-binding model, for example periodic waveguide
arrays that provide a classical, electric field analog of a quantum
system with exponential decay \cite{Lon2}. 

%
We thank A. del Campo, D. Gu\'ery-Odelin, J. Martorell, D. Sprung,
and E. Sherman for discussions.
We acknowledge 
funding by the Basque Government (Grants No. IT472-10 and BFI08.151)
and Ministerio de 
Ciencia e Innovaci\'on (FIS2009-12773-C02-01).

\end{document}